\documentclass[aps,prl,twocolumn,superscriptaddress,amsmath,amssymb,longbibliography, floatfix]{revtex4-2}

\usepackage[english]{babel}
\usepackage[utf8]{inputenc}

\usepackage{hyperref}
\usepackage{graphicx}
\usepackage{array}
\usepackage{amsmath,amsfonts,amssymb,bm}
\usepackage{braket}
\usepackage{color}
\usepackage{calc}
\usepackage[normalem]{ulem}
\usepackage{bibentry}
\usepackage{pbox}

\begin{document}

\title{Momentum-Locked Edge Excitons in 2D Molecular Crystals}

\author{Sidhartha Nayak}
\affiliation{Institute of Physics,\unpenalty~Czech Academy of Sciences,\unpenalty~Cukrovarnick\'{a} 10,\unpenalty~Prague,\unpenalty~16200,\unpenalty~Czech Republic}

\author{Alex Boeglin}
\affiliation{Universit\'e de
Strasbourg,\unpenalty~CNRS,\unpenalty~IPCMS,\unpenalty~UMR 7504,\unpenalty~F-67000 Strasbourg,\unpenalty~France}

\author{Quentin Stein}
\affiliation{Universit\'e de
Strasbourg,\unpenalty~CNRS,\unpenalty~IPCMS,\unpenalty~UMR 7504,\unpenalty~F-67000 Strasbourg,\unpenalty~France}

\author{Katharina Kaiser}
\affiliation{Universit\'e de
Strasbourg,\unpenalty~CNRS,\unpenalty~IPCMS,\unpenalty~UMR 7504,\unpenalty~F-67000 Strasbourg,\unpenalty~France}
\affiliation{IV. Physical Institute -- Solids and Nanostructures,\unpenalty~Georg-August-Universit\"{a}t G\"{o}ttingen, G\"{o}ttingen,\unpenalty~37077,\unpenalty~Germany}

\author{Guillaume Schull}
\affiliation{Universit\'e de
Strasbourg,\unpenalty~CNRS,\unpenalty~IPCMS,\unpenalty~UMR 7504,\unpenalty~F-67000 Strasbourg,\unpenalty~France}

\author{Thomas Frederiksen}
\affiliation{Donostia International Physics Center,\unpenalty~Donostia-San Sebasti\'{a}n 20018,\unpenalty~Spain.}
\affiliation{Ikerbasque, Basque Foundation for Science, E-48013 Bilbao, Spain}

\author{Tom\'a\v{s} Neuman}
\affiliation{Institute of Physics,\unpenalty~Czech Academy of Sciences,\unpenalty~Cukrovarnick\'{a} 10,\unpenalty~Prague,\unpenalty~16200,\unpenalty~Czech Republic}

\date{\today}

\begin{abstract}
In molecular crystals, molecular excitations can hybridize into collective Frenkel excitons. Here we show that doubly degenerate excitations in planar two-dimensional molecular crystals generate excitonic bands with nontrivial momentum-resolved topological characteristics and edge states appearing only for selected crystallographic terminations. Using a model inspired by aggregates of zinc phthalocyanine molecules deposited on insulating surfaces, we demonstrate that the existence of edge excitons depends sensitively on the crystallographic orientation of the boundary and is governed by the Zak phase of the underlying bulk bands. The resulting edge states exhibit momentum-locked elliptical polarization textures originating from the internal structure of the molecular excitations, potentially enabling directional optical coupling and excitation transport. Our results establish a route toward topological and polarization-textured excitonic states in two-dimensional molecular materials.
\newline

\end{abstract}

\maketitle

\begin{figure}[ht]
    \centering
    \includegraphics[width=\linewidth]{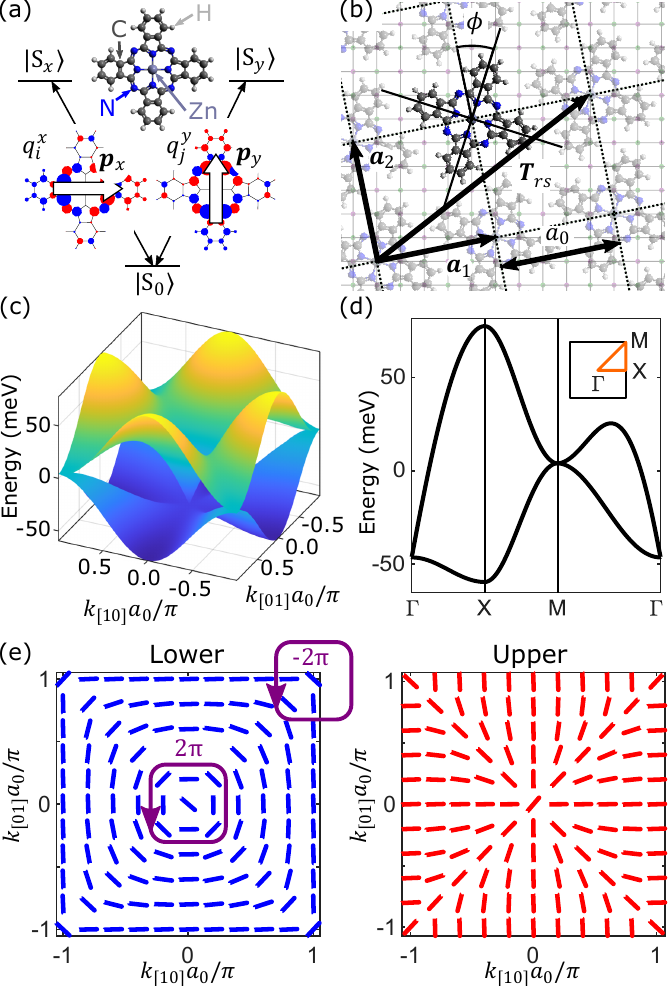}
    \caption{(a) The monomer, zinc phthalocyanine (ZnPc) molecule (center top), electronic structure is modeled using three states: the ground state $\vert {\rm S}_0\rangle$ and two degenerate excited states $\vert {\rm S}_x\rangle$ and $\vert {\rm S}_y\rangle$. The $\vert {\rm S}_0\rangle$ $\to$ $\vert {\rm S}_\alpha \rangle$, with $\alpha\in\{x,y\}$, transitions are associated with transition charge densities, here represented by the effective charges $q_i^\alpha$ localized on atom $i$. The distribution of effective charges defines the transition dipoles ${\bf p}_\alpha $ shown as white arrows. (b) Geometry of the ZnPc lattice including the definitions of the primitive vectors ${\bf a}_1$ and ${\bf a}_2$, the lattice constant $a_0$, a generic lattice vector ${\bf T}_{rs}$, and the angle $\phi$ of rotation of the molecule in the unit cell. (c) The band structure of the infinite 2D lattice, offset with respect to the excitation energy $E_0\approx 1.85$ eV of the isolated molecule. (d) The band structure plotted along the path ${\rm \Gamma}- {\rm X}-{\rm M}-{\rm \Gamma}$. The schematic of the Brillouin zone is provided in the inset. (e) Polarization texture shown as polarization ellipses (degenerating here into line segments) of the lower (blue) and upper (red) exciton branch, respectively. We indicate possible integration paths (purple contours) for the evaluation of Berry phase around $\Gamma$ and M, both resulting in $\gamma=0$ (mod $2\pi$).}
\label{fig:model}
\end{figure}

When molecules are brought together into a crystal, interactions between them give rise to collective Frenkel excitons \cite{Frenkel1931,Davydov1964,Hestand2017,Hestand2018}. In extended systems, they form excitonic bands whose properties are driven by the electronic structure of the monomer, but also by the crystal geometry. Frenkel excitons have been widely studied for their contribution to long-range energy transfer and their characteristic signatures in optical spectra and near-field optical microscopy images \cite{Chuang2019, Zheng2019, Kim2023, Juergensen2023, Cheng2025, Hasan2025}. However, less is known about their coherent behavior and particularly about the topological aspects of their band structure under well-controlled conditions \cite{Thompson2025, Zhou2014, Gao2018}. Questions such as the possibility of long range energy transfer or the possible existence of boundary states in certain non-trivial topologies arise in this context.
Here we consider a realistic system of zinc phthalocyanine (ZnPc) molecules deposited on the (100) surface of sodium chloride (NaCl), where they form two-dimensional (2D) molecular crystals. These aggregates have been experimentally demonstrated and their excitonic properties can be probed by light scanning tunneling microscopy (STML), where coherent intermolecular excitonic interactions have already been observed in small clusters \cite{Zhang2016, Luo2019, Dolezal2021b, Jing2026}. Recent experiments on larger two-dimensional aggregates further revealed signatures consistent with both bulk and edge excitonic states \cite{Stein2026}, motivating a theoretical investigation of the momentum-resolved topology and boundary excitons in these systems. 

To model Frenkel excitons in 2D aggregates of ZnPc molecules we consider a Hamiltonian model based on several assumptions. We describe the molecular monomer as a three-level system composed of its electronic ground state $\vert {\rm S}_0\rangle$ and two degenerate excited states $\vert {\rm S}_x\rangle\equiv\vert x\rangle$ and $\vert {\rm S}_y\rangle\equiv\vert y\rangle$, as shown in Fig.\,\ref{fig:model}a. Each molecule thus contains two degenerate transitions $\vert {\rm S}_0\rangle$ $\to$ $\vert {\rm S}_x\rangle$ and $\vert {\rm S}_0\rangle$ $\to$ $\vert {\rm S}_y\rangle$, that are accompanied by their corresponding transition charge densities $\rho_x$ and $\rho_y$ (defining the transition dipoles ${\bf p}_x$ and ${\bf p}_y$). 
The interaction between monomers $m$ and $n$ are introduced by the unscreened Coulomb interaction $V_{mn}^{\alpha\beta}$ ($\alpha, \beta\in \{x,y\}$). 
We evaluate the Coulomb integral using the approximation where the transition densities $\rho_\alpha  ({\bf r})$ are expressed in terms of effective charges $q_i^\alpha$ at each atom $i$ as
$V_{mn}^{\alpha\beta} \approx \sum_{ij} {q_i^\alpha q_j^\beta}/{|{\bf r}_i+{\bf R}_m-({\bf r}_j+{\bf R}_n)|}$,
where we assume that the positions ${\bf r}_i$ of the equivalent charges are centered around the coordinate origin and ${\bf R}_m$ is the position of the center of monomer $m$ (see Fig.\,\ref{fig:model}b). 

We next restrict the number of excitations in the system to one. In this case, the Frenkel Hamiltonian of the aggregate becomes (subtracting the monomer onsite energy $E_0\approx 1.85$eV) \cite{Davydov1964}
\begin{align}
    \hat{H}=\sum_{\substack{m\neq n \\ \alpha\beta}}V_{mn}^{\alpha\beta}\vert n,\beta\rangle\langle m,\alpha\vert,\label{eq:hamiltonian}
\end{align}
where $\vert n,\beta\rangle$ is the excited state $\vert {\rm S}_\beta  \rangle$ of monomer $n$ while all other molecules are in their ground state. Due to the degeneracy of the monomer excitations, this Hamiltonian introduces an effective pseudospin encoded in the polarization state of the monomer excitations. In this work, we do not include vibronic effects \cite{Hestand2018} as the molecule of ZnPc features negligible vibronic activity due to its overall rigidity \cite{Doppagne2017}. 

To elucidate the properties of the 2D excitonic material, we calculate its band structure for the square lattice shown in Fig.\,\ref{fig:model}b, with lattice vectors ${\bf T}_{rs}=r{\bf a}_1+s{\bf a}_2$ and a single molecule per unit cell rotated by an angle $\phi$ with respect to the lattice axes. This geometry reflects low-temperature experiments on ZnPc molecules adsorbed on the NaCl(100) surface \cite{Zhang2016}, where $\phi\approx29^\circ$ and $a_0\approx1.44$ nm \cite{Zhang2016, Stein2026}. While the molecular rotation reduces the system symmetry to C$_4$ and slightly modifies the excitonic dispersion, the qualitative features discussed below arise from the twofold degeneracy of the molecular excitation and the tensorial nature of the intermolecular Coulomb coupling and are therefore not specific to this particular adsorption geometry.
Moreover, the molecular configuration minimizes direct orbital overlaps between nearest neighbors \cite{Zhao2019, Fornari2016} and renders the intermolecular coupling predominantly long-range Coulombic.

To describe the excitonic properties, we build a lattice Hamiltonian using Eq.\,\eqref{eq:hamiltonian} and transform it into the reciprocal space by projecting it into the Bloch-wave basis
$\vert {\bf k}, \alpha\rangle=\frac{1}{\sqrt{N_{\rm m}}}\sum_m e^{i{\bf k}\cdot {\bf r}_m}\vert m, \alpha\rangle$,
where $N_{\rm m}$ is the number of monomers that diverges in the limit of infinite crystal. The Hamiltonian becomes
\begin{align}
    \hat{\tilde{H}}({\bf k})=\sum_{\alpha\beta} \tilde{H}_{\alpha\beta}({\bf k})\vert {\bf k}, \alpha\rangle\langle {\bf k}, \beta \vert,\label{eq:hamiltonianK}
\end{align}
with $\tilde{H}_{\alpha\beta}({\bf k})=\sum_{{\bf T}_{rs}\neq 0} V_{{\bf T}_{rs}}^{\alpha\beta}e^{i{\bf k}\cdot {\bf T}_{rs}}$,
and 
$V^{\alpha\beta}_{{\bf T}_{rs}}=\sum_{ij}{q_i^\alpha q_j^\beta}/{|{\bf r}_i-{\bf r}_j-{\bf T}_{rs}|}$.
To gain insight into the structure of the Frenkel excitons determined by this square lattice we solve the Hamiltonian parametrically for a grid of ${\bf k}$ vectors and calculate the excitonic band structure shown in Fig.\,\ref{fig:model}c. For further convenience we show the ${\rm \Gamma}-{\rm X}-{\rm M}-{\rm \Gamma}$ cut through the Brillouin zone (BZ) in Fig.\,\ref{fig:model}d, sketching the BZ geometry in the inset. 
Linearly dispersing bands are found around the degeneracy at ${\rm \Gamma}$ and a quadratic touching at M. 
At the X point of the BZ we find a maximally opened gap between the two excitonic bands. 

The energy bands can be associated with the underlying exciton wave functions. In Fig.\,\ref{fig:model}e we represent the real-space polarization (not to be confused with the pseudospin which reflects the polarization state) of the underlying Bloch functions. In general, the polarization is indicated by ellipses that represent the trajectory of the head of the polarization vector within a molecular monomer during its temporal oscillation. As the Bloch functions are linearly polarized (both polarization components oscillate in phase), the polarization is represented by straight lines forming distinct textures for the lower (blue) and upper (red) energy band throughout the BZ. The overall pattern of these lines is centered around the ${\rm \Gamma}$ point and forms a circular pattern for the lower band and a radial pattern for the upper band. At ${\rm \Gamma}$ and M the polarization is not well defined due to the degeneracy of the energy bands. 

The degeneracy at ${\rm \Gamma}$ is actually a quadratic touching with double winding of the pseudospin, not a single Dirac cone. This can be directly seen in the polarization textures in Fig.\,\ref{fig:model}e as the pseudospin (polarization state - in this case directly mapping to the orientation of the polarization line) is the same exactly twice when looping around $\Gamma$.
As we detail in the SM \cite{supplementary}, a quadratic touching is actually expected at ${\rm \Gamma}$ from symmetry. However, due to the additional $1/|{\bf k}|$ factor arising from the long-range character of the interaction $V_{mn}^{\alpha\beta}$ \cite{Ford1984}, a linear dispersion is observed. The long-range nature of the Coulomb interaction therefore produces a linearly dispersing quadratic touching, highlighting that linear dispersion alone is insufficient to identify Dirac topology when long-range interactions are involved. This is also confirmed by calculating the Berry phase \cite{Berry1984, Zak1989, Xiao2010, Delplace2011}  $\gamma$ along a path enclosing $\Gamma$, yielding $\gamma=0$ (mod $2\pi$), differing from $\gamma=\pi$ expected for a genuine Dirac cone.  

Dirac cones, however, can be obtained from the quadratic touchings at $\Gamma$ and M by breaking the crystal symmetry. One possibility consists in applying a perturbation such as $\hat{H}_{\rm p}=\sum_m\frac{\delta}{2}\left ( \vert m, x \rangle\langle  m,x \vert - \vert m,y \rangle\langle  m,y \vert\right )$ with the on-site energy $\delta$, splitting the $\vert {\rm S}_x\rangle$-$\vert {\rm S}_y\rangle$ degeneracy \cite{supplementary}, which could be achieved by Stark-shifting the molecular excitation by external electric fields \cite{Arrieta2026} or by choosing a different monomer with near-degenerate excited states\cite{Roslawska2026}. Such perturbation splits both degeneracies at ${\rm \Gamma}$ and M, each into a pair of conical crossings carrying $\gamma=\pi$ (mod $2\pi$), indicating that all four conical features are of Dirac character. Note that there are other theoretical possibilities how to lift this degeneracy. For example, one could design a rectangular lattice by manipulating the inter-molecular spacing. 
 
\begin{figure}[t]
    \centering
    \includegraphics[width=\linewidth]{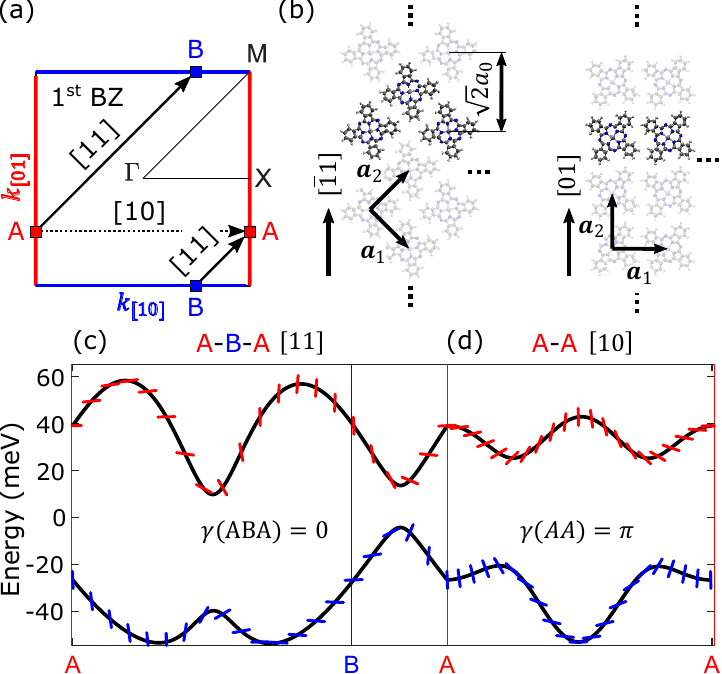}
    \caption{(a) The first BZ marking the significant points and two points on the zone boundary: A and B. The edges are color coded to reflect the toroidal topology of the BZ. Two integration paths proceeding A$\to$A and A$\to$B$\to$A marked according to their crystallographic directions as [10] and [11], respectively. (b) Schematic depiction of edges cut along the [$\bar 1$1] and [01] directions. (c,d) Cuts through the band structure along the paths (c) A$\to$B$\to$A ([11]) and (d) A$\to$A ([10]). The calculated Zak phase $\gamma$ (identical for both bands) is shown in the inset. The bars overlaying the bands mark the polarization of the respective wave functions as in Fig.\,\ref{fig:model}e.}
\label{fig:edge}
\end{figure}

Beyond revealing the character of degeneracies at $\Gamma$ and M, the polarization texture in Fig.\,\ref{fig:model}e also suggests that, when edges of certain crystallographic directions are introduced, they may host edge-localized states. To show this, we analyze the bulk band structure using the Zak phase \cite{Su1980, Zak1989}, i.e., the band-resolved Berry phase accumulated along closed loops in reciprocal space. For a given edge orientation, the integration path is chosen perpendicular to the edge direction while keeping the projected crystal momentum fixed \cite{Delplace2011}. Non-trivial Zak phase ($\gamma=\pi$ mod $2\pi$) indicates that when cut along the corresponding edge, edge-localized states appear. Trivial phase ($\gamma=0$ mod $2\pi$) indicates the lack thereof. We calculate the Zak phase along two significant directions in the BZ, [10] and [11], indicated in Fig.\,\ref{fig:edge}a as dotted and full arrows traversing the BZ, both starting and finishing at point A. These directions in the BZ correspond to edges cut in real space along the crystallographic directions [01] and [$\bar{1}$1], respectively, as shown in Fig.\,\ref{fig:edge}b. We color coded the boundaries of the BZ to stress its toroidal character; edges of the same color are identical and both paths [10] (A$-$A) and [11] (A$-$B$-$A) are therefore closed loops. In Fig.\,\ref{fig:edge}c,d we show the cross section of the energy band structure along the respective paths. Both cross sections have the character of gapped 1D bands and therefore allow us to unambiguously evaluate the band-resolved Zak phase. Along the [10] (A$-$A) cut both the upper and lower bands are nontrivial, while, in contrast, along the [11] (A$-$B$-$A) path the bands are trivial. We note that the Zak phase does not depend on the precise position of point A as long as the cut along the respective directions does not intersect any of the degeneracy points of the band structure. This follows from the trivial Berry phase obtained by integrating along closed loops around both ${\rm \Gamma}$ and M (purple loops in Fig.\,\ref{fig:edge}a). 

We can build a more intuitive understanding of the Zak phase by directly inspecting the polarization textures in Fig.\,\ref{fig:model}e. For the present system, $\gamma$ is given by the accumulated rotation of the polarization lines (pseudospin winding) along the integration path. Consequently, the existence of edge states for an arbitrary crystallographic orientation can be inferred graphically by following the evolution of the polarization texture along the corresponding path in reciprocal space. As examples, the A$-$A path accumulates a rotation of $\gamma=\pi$, whereas the A$-$B and B$-$A segments accumulate rotations of $-\pi/2$ and $\pi/2$, respectively, and therefore cancel, $\gamma=0$. 
This construction can be generalized to an arbitrary crystallographic cut $[mn]$ chosen not to intersect any degeneracy point. 
We show in the SM \cite{supplementary} that $\gamma=N_{\pi}\pi$ (mod $2\pi$), with $N_{\pi}=||m|-|n||$.   

\begin{figure}[t]
    \centering
    \includegraphics[width=0.95\linewidth]{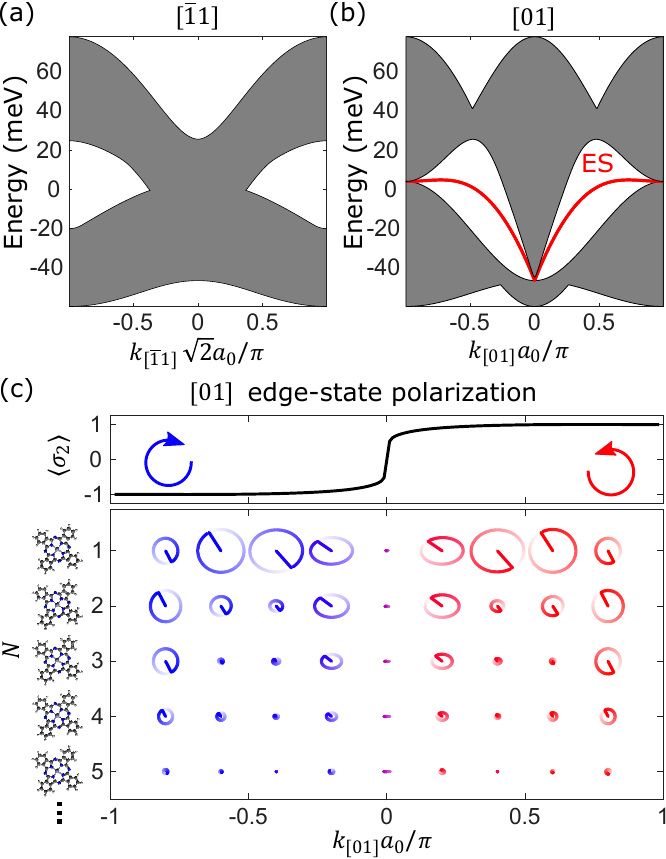}
    \caption{(a) The band structure of a wide ribbon with [$\bar 1$1] edges and (b) the band structure of a wide ribbon with the [01] edges. The states emerging from the bulk band structure (gray shaded areas), red lines marks the edge states (ES). (c) ES polarization of the edge molecules for different $k_{[01]}$ expressed as an expectation value of the $\sigma_2$ Pauli matrix, and polarization ellipses of five molecular excitations forming the edge states for selected values of $k_{[01]}$ along the [01] edge. The color coding marks the component of the right (red) and left (blue) handed state. The bars inside the ellipses indicate the instantaneous orientation of their oscillating polarization vector and the ellipse shading further visualizes its sense of rotation.}
\label{fig:polar}
\end{figure}

Based on the analysis of the Zak phase, we therefore anticipate that the edges in the $\langle 01 \rangle$ family host localized states but the $\langle\bar{1}1\rangle$ ones do not. We demonstrate this computationally by designing 1D Hamiltonians of wide ribbons defined by the appropriate edges. 
In Fig.\,\ref{fig:edge}b we also highlight the unit cell of the 1D Hamiltonians of the respective ribbons. In the case of the $[\bar 1 1]$ ribbon, the unit cell consists of a chain of ZnPc dimers, while in the case of the [01] ribbon, the unit cell consists of a chain of ZnPc monomers. 
We show the 1D band structure calculated for the [$\bar 1$1] edge in Fig.\,\ref{fig:polar}a. It contains a (discretized) continuum of states (gray shaded area) that emerges as a projection of the bulk band structure onto the edge reciprocal vector. In contrast, the 1D band structure calculated for the [01] ribbon, shown in Fig.\,\ref{fig:polar}b, features additional states (red lines) in the projected band gap of the bulk states (gray shaded area). This new part of the spectrum is a signature of the anticipated edge states (ES). These results fully confirm the analysis based on the momentum-resolved 2D-band topology.   

The nature of the states localized at the [01] edge is revealed by their wave functions. Figure\,\ref{fig:polar}c shows polarization ellipses for selected periodic parts of the ribbon Bloch functions, analogous to Fig.\,\ref{fig:model}e. Here, the horizontal axis corresponds to the crystal momentum along the ribbon, while the vertical one denotes the molecular position $N$ within the semi-infinite unit cell, measured from the edge ($N=1$), as illustrated schematically on the left. Each column thus represents a different wave function for a different $k_y$. The ellipses are color coded according to their handedness: red indicates right-handed (positive sense of rotation \cite{handedness_note}) and blue left-handed polarization. The bar inside the ellipse represents the instantaneous orientation of the oscillating vector of polarization. The degree of circular polarization is also quantified for the first edge molecule as an expectation value of the Pauli matrix $\sigma_2$ (black curve in Fig.\,\ref{fig:polar}c). 

Several features are noteworthy. At the center and boundaries of the 1D BZ, the states are delocalized and therefore do not constitute true edge states. This reflects the closing of the projected bulk band gap at the corresponding bulk degeneracies (Fig.\,\ref{fig:polar}b). Moving away from these points, the states become exponentially localized at the edge, with the decay length decreasing as the projected gap opens. Apart from the sense of polarization, the wave functions for $\pm |k_{[01]}|$ look almost identical, despite the lower C$_4$ symmetry of the system that does not guarantee this property. The locking of polarization and momentum is another hallmark of the edge state. States with $k_{[01]}<0$ are left-handed, whereas those with $k_{[01]}>0$ are right-handed. This behavior is analogous to momentum-polarization locking in topological photonics \cite{Rodriguez-Fortuno2013, Mitsch2014, Xiong2022, Hafezi2011, Ozawa2019} and, more generally, to spin-momentum locking in topological insulators \cite{Hsieh2008, Chen2009b, Moore2010}. It enables selective excitation of counter-propagating edge states through the polarization of external optical probes and may facilitate unidirectional energy transfer from emitters with circularly polarized transition dipoles \cite{Mitsch2014}. As the BZ center is approached, the polarization becomes increasingly eccentric and ultimately evolves into a linearly polarized state where the projected gap closes.

We have shown that molecular crystals composed of molecules with doubly degenerate excitations support excitonic edge states whose existence depends on the crystallographic orientation of the boundary. These states exhibit momentum-locked polarization textures arising from the internal two-component structure of the excitons. Together with recent STML observations of bulk and edge excitonic modes \cite{Stein2026}, our results establish molecular aggregates as a platform where momentum-resolved topology gives rise to polarization-textured excitonic edge states.

\section{Acknowledgement}
G.S. acknowledges funding from the European Research Council (ERC) under the European Union's Horizon 2020 research and innovation program (grant agreement No 101142565) and from the Agence Nationale de la Recherche and the Swiss National Science Foundation under the common grant ANR-SNF GRAAL (ANR-23-CE09-0030). This work of the Interdisciplinary Thematic Institute QMat, as part of the ITI 2021 2028 program of the University of Strasbourg, CNRS and Inserm, was supported by IdEx Unistra (ANR 10 IDEX 0002), and by SFRI STRAT'US project (ANR 20 SFRI 0012) and EUR QMAT ANR-17-EURE-0024 under the framework of the French Investments for the Future Program.
S.N. and T.N. acknowledge the Lumina Quaeruntur fellowship of the Czech Academy of Sciences and QUANTERA project MOLAR with reference PCI2024-153449. Computational resources were provided by the e-INFRA CZ project (ID:90254), supported by the Ministry of Education, Youth and Sports of the Czech Republic.
T.F. acknowledges support from the Spanish MCIN / AEI / 10.13039/501100011033 (PID2023-146694NB-I00, GRAFIQ).


%

\clearpage

\renewcommand{\thefigure}{S\arabic{figure}}
\renewcommand{\theequation}{S\arabic{equation}}

\setcounter{equation}{0}
\setcounter{figure}{0}

\onecolumngrid
\begin{center}
{\large\bfseries Supplementary Material for ``Momentum-Locked Edge Excitons in 2D Molecular Crystal''}
\end{center}
\twocolumngrid
\section{TDDFT calculations and effective charges}

We evaluate the interaction between the molecules using equivalent charges $q_i^\alpha$ ($\alpha\in\{x,y\}$) localized at the atomic centers $i$, approximating the transition density $\rho_\alpha\approx \sum_i q_i^\alpha\delta({\bf r}-{\bf r}_i)$, as
$V_{mn}^{\alpha\beta} \approx \sum_{ij} {q_i^\alpha q_j^\beta}/{|{\bf r}_i+{\bf R}_m-({\bf r}_j+{\bf R}_n)|}$,
where we assume that the positions ${\bf r}_i$ of the equivalent charges (atomic centers) are referenced to the central atom of the molecule, and ${\bf R}_m$ is the position of the center of monomer $m$. This approximates the integral $V_{mn}^{\alpha\beta} = \iint {\rho_\alpha  ({\bf r})\rho_\beta  ({\bf r}')}/{|{\bf r}-{\bf r}'+{\bf R}_m-{\bf R}_n|}{\rm d}^3{\bf r}{\rm d}^3{\bf r}'$, where the transition densities $\rho_\alpha  ({\bf r})$ and $\rho_\beta  ({\bf r}')$ are centered around ${\bf R}=0$.

To obtain $q_i^\alpha$ we first extract the transition density $\rho_\alpha$ from linear-response time-dependent density-functional calculations performed in Gaussian 16, revision C.01 \cite{g16}. We employ the B3LYP hybrid functional \cite{Becke1993, Lee1988, Becke1993a} and the Pople 6-31G* basis set. The molecular geometry is enforced to have the D$_{4\rm h}$ symmetry and is optimized in the electronic ground state using the same functional and basis set.
The transition densities were extracted using the Multiwfn 3.8 tool \cite{Lu2012}, which was also used to convert the densities into the charges from electrostatic potentials using a grid-based method (CHELPG) that are localized on the molecular atomic centers. This method is implemented in the LIBRETA library \cite{Zhang2021b} utilized by Multiwfn. CHELPG are found such that their electrostatic potential approximates the one of the full three-dimensional density on a grid set up in the vicinity of the molecule. We use a grid defined by the maximal distance from the molecular atomic centers of $20$\,Bohr and grid spacing of $0.8$\,Bohr. We use the default minimal distances from the atoms used by the LIBRETA library (1.450\,{\AA} for H, 1.5\,{\AA} for C, and 1.7\,{\AA} for N), and for Zn the atomic radius of 1.1513\,{\AA}. 

\section{Evaluation of Berry (Zak) phase}

In practice, implement the evaluation of the Berry phase on an equally spaced $201\times201$ square ${\bf k}$-point grid covering the Brillouin zone. The Berry phase is calculated as
\begin{align}
    \gamma\approx -{\rm Im}\left\{\ln\prod_{i\in \mathcal C} \langle u({\bf k}_i)\vert u({\bf k}_{i+1})\rangle\right\},
\end{align}
where $\mathcal C$ is the considered contour, the first and the last point coincide, and $\vert u({\bf k}_i) \rangle$ is the periodic part of the Bloch function of the band in question. When evaluating closed loops around $\Gamma$ and M, we choose rectangular integration paths (although the geometric shape is unimportant), while the Zak phase for the momentum-resolved topology is evaluated along linear cuts indicated in the main text. In the latter case, it is important to ensure that the last point of integration is numerically chosen to be the starting point of the integration to avoid numerical issues. 

\section{Degeneracy around \texorpdfstring{$\Gamma$}{Gamma}}

The Hamiltonian in Eq.\,\eqref{eq:hamiltonianK} of the main text is expressed as a lattice sum in real space. In order to analyze its behavior around ${\rm \Gamma}$, we transform it into reciprocal space using the identity
\begin{align}
    \sum_{\bf T} V^{\alpha\beta}({\bf T})e^{i{\bf k}\cdot {\bf T}}= \frac{1}{A_{\rm cell}}\sum_{{\bf G}} \tilde V^{\alpha\beta}({\bf k}+{\bf G}),
\end{align}
where $\tilde V^{\alpha\beta}({\bf q})=\int \, V^{\alpha\beta}({\bf R})e^{i{\bf q}\cdot {\bf R}}{\rm d}^2{\bf R}$ is the Fourier transform of $V^{\alpha\beta}({\bf R})$, ${\bf G}$ are the reciprocal lattice vectors, and $A_{\rm cell}=a_0^2$ is the area of the unit cell. We have also dropped the indices $r$ and $s$ of the lattice vectors ${\bf T}_{rs}$ for brevity. 
We next evaluate $\tilde V^{\alpha\beta}({\bf q})$
\begin{align}
    \tilde V^{\alpha\beta}({\bf q})&=\sum_{mn} \int  q^\alpha_m q^\beta_n \frac{e^{i{\bf q}\cdot{\bf R}}}{|{\bf R}+{\bf r}_n-{\bf r}_m|}{\rm d}^2{\bf R}\nonumber\\
    &=\int \frac{e^{i{\bf q}\cdot{\bf X}}}{|{\bf X}|}  {\rm d}^2{\bf X}\sum_{m} q_m^\alpha e^{i{\bf q}\cdot{\bf r}_m}\sum_n q_n^\beta e^{-i{\bf q}\cdot{\bf r}_n} \nonumber\\
    &= \frac{2\pi}{|{\bf q}|}\sum_{m} q_m^\alpha e^{i{\bf q}\cdot{\bf r}_m}\sum_n q_n^\beta e^{-i{\bf q}\cdot{\bf r}_n},
\end{align}
where in the second step we used the substitution ${\bf X}={\bf R}+{\bf r}_n-{\bf r}_n$.
We next expand the ${\bf q}$-dependent exponential functions, noticing that due to charge neutrality of the transition densities, the constant term vanishes and we only retain the term linear in ${\bf k}$. 
The reciprocal-space Hamiltonian becomes
\begin{align}
    \tilde H_{\alpha\beta}({\bf k}) &= \frac{2\pi}{a_0^2}\sum_{\bf G} \frac{1}{|{\bf k}+{\bf G}|} \sigma^\alpha({\bf k}+{\bf G})\sigma^\beta(-{\bf k}-{\bf G})\nonumber\\
    &-V^{\alpha\beta}(0),
\end{align}
with $\sigma^\alpha({\bf q})=\sum_{m} q_m^\alpha e^{i{\bf q}\cdot{\bf r}_m}$, and we subtracted the self-interaction term $V^{\alpha\beta}(0)$.
To analyze the behavior around ${\bf k}=0$ we can further expand the Hamiltonian, noticing that the ${\bf G}=0$ term is the dominant contribution containing a $1/|{\bf k}|$ divergence:
\begin{align}
    \tilde H^{\rm dom}_{\alpha\beta}({\bf k}) &= \frac{2\pi p^2}{a_0^2}\frac{k_\alpha k_\beta}{|{\bf k}|},\label{eq:dom}
\end{align}
where we have only retained the dominant term, dropped constant corrections, and introduced the magnitude $d$ of the transition dipole moment $p=|{\bf p}_\alpha|$, with ${\bf p}_\alpha=\sum_m q_m^\alpha{\bf r}_m$. The Hamiltonian in Eq.\,\eqref{eq:dom} of the main text can be rewritten using Pauli matrices 
\begin{align}
    \tilde H^{\rm dom}_{\alpha\beta}({\bf k}) &=\frac{\pi p^2}{a_0^2}\left[\frac{k_\alpha ^2+k_\beta ^2}{|{\bf k}|}\sigma_0+\frac{2k_\alpha k_\beta }{|{\bf k}|}\sigma_1+\frac{k_\alpha ^2-k_\beta ^2}{|{\bf k}|}\sigma_3\right]\nonumber\\
    &=\frac{\pi p^2}{a_0^2}\left[ {k_\rho}\sigma_0+k_\rho\sin2\varphi\,\sigma_1+k_\rho\cos 2\varphi\,\sigma_3   \right],\label{eq:hampauli}
\end{align}
with $k_\alpha=k_\rho\cos\varphi$, $k_\beta=k_\rho\sin\varphi$, and $|{\bf k}|=k_\rho$. The Hamiltonian in Eq.\,\eqref{eq:hampauli} of  the main text has a linear dispersion, but features a winding of 2 around ${\rm \Gamma}$.
We also note that this linear behavior is only possible around $\Gamma$ due to the ${\bf G}=0$ term. Such a term does not exist for other points in the first BZ; that is why an ordinary quadratic touching is observed at M. 

To see that the cone-like structure at $\Gamma$ is a quadratic touching and not a Dirac cone, we also numerically evaluate the Berry phase
$\gamma = i\oint_{C_{\rm \Gamma}} \langle u_i({\bf k})\vert \nabla_{\bf k}u_i({\bf k})\rangle \cdot {\rm d}{\bf k}$ for each energy band of the full Hamiltonian in Eq.\,\eqref{eq:hamiltonianK} on a loop enclosing $\Gamma$.
Here $u_i({\bf k})$ is the Bloch function of band $i$ and $\nabla_{\bf k}$ is the gradient in reciprocal space. The presence of a genuine Dirac cone \cite{Mikitik1999, Castro2009} requires $\gamma=\pi$ (mod $2\pi$). However, we obtain $\gamma=0$ (mod $2\pi$), consistently with our analytical treatment.

\section{Lifting the monomer degeneracy}

\begin{figure}[t]
    \centering
    \includegraphics[width=\linewidth]{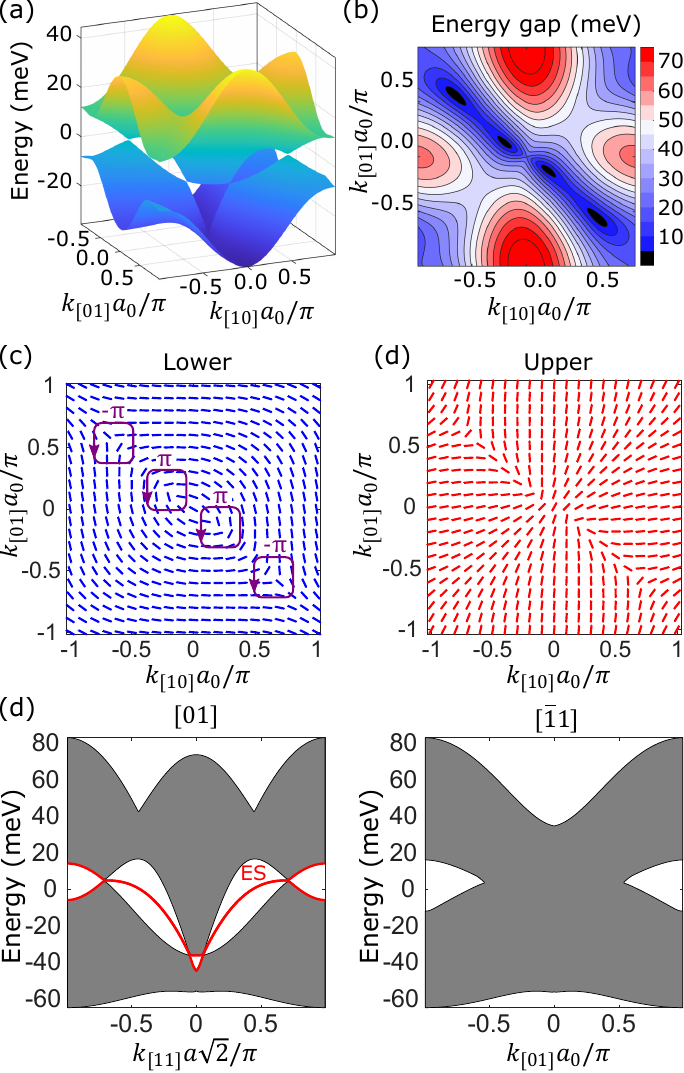}
    \caption{(a) The band structure of the infinite 2D lattice perturbed by $\hat{H}_{\rm p}$, with $\delta=-20$\,meV. (b) A contour plot of the energy of the band gap. Black regions mark the approximate positions of the crossings.  (c,d) Polarization texture, shown as polarization ellipses (degenerating here into line segments), of the (c) lower (blue) and (d) upper (red) exciton branch, respectively. Berry phase contours (purple) and the corresponding values are indicated in (c). (d) Band structure of the edges created by introducing a cut along the [01] and [$\overline{1}1$] direction, respectively, as descrived in the main text. }
\label{fig:splitting}
\end{figure}

As discussed in the main text, when a perturbation is applied, such as $\hat{H}_{\rm p}=\sum_m\frac{\delta}{2}\left ( \vert m, x \rangle\langle  m,x \vert - \vert m,y \rangle\langle  m,y \vert\right )$ with the on-site energy $\delta$, splitting the $\vert {\rm S}_x\rangle$-$\vert {\rm S}_y\rangle$ degeneracy, the degeneracies at $\Gamma$ and M split into pairs of Dirac cones. The full band structure is shown in Fig.\,\ref{fig:splitting}a. For clarity, we also plot the magnitude of the gap between the upper and lower excitonic branches in Fig.\,\ref{fig:splitting}b. Four degeneracy points marking Dirac cones are located in the four black areas where the gap closes. The modification of the band structure is also reflected in the polarization texture of the lower (Fig.\,\ref{fig:splitting}c) and upper (Fig.\,\ref{fig:splitting}d) branch. The polarization texture is represented by polarization ellipses that have the form of straight lines as the polarization is linear everywhere in the BZ. The four degeneracies represent the points where the polarization is arbitrary (more precisely, an arbitrary pair of orthogonal polarization vectors).       

Following the intuitive understanding of the Berry and Zak phases described in the main text, we can inspect the polarization textures along closed loops around the degeneracies, revealing a single pseudospin winding in all cases. Similarly, the Zak phases can be read out when considering linear integration paths through the BZ, showing that the phases depend on the relative position of the integration paths with respect to the four Dirac cones and the angle with which the path traverses the BZ. This means that the [01] edge states now exist only when the integration paths pass for edge momenta lying between the Dirac cone emerging from the original degeneracy at $\Gamma$ and the Dirac cone originating from the original degeneracy at M. This is demonstrated in Fig.\ref{fig:splitting}d, where the 1D band structure of both the [01] and [$\overline{1}1$] edges is shown including the perturbed Hamiltonian $\hat{H}_{\rm p}$.

\section{Edge band structure}

In the main text we present band structures calculated for the semi-infinite system with two different edge terminations. Here we outline the procedure to obtain these band structures from the Hamiltonian in Eq.\,\eqref{eq:hamiltonian}. For an arbitrary edge [$mn$] one can define a corresponding supercell of the fully 2D problem by the vectors [$n\bar{m}$] and [$mn$]. This supercell contains $m^2+n^2$ molecules, each carrying two internal polarizations, resulting in $2(m^2+n^2)$ states per supercell. Numerically, we construct the supercell Hamiltonian on a finite grid defined by the two translation vectors and then perform a 1D Fourier transform in the direction of the edge. For the [01] edge, the supercell is identical with the primitive cell and we thus obtain the edge Hamiltonian directly from the Hamiltonian of the 2D crystal by only restricting the Fourier transform to the [01] direction. In practice we use a grid of $201\times 201$ molecules, which is sufficient to suppress the interaction of the states localized on the opposing edges and reproduce the projected bands of the bulk crystal. For the $\langle 11\rangle$ edge, we construct the Hamiltonian on a $141\times 141$ grid of molecular dimers, again ensuring sufficient length of the finite-sized chain so as to reproduce the discretized projected band structure of the bulk crystal.

Notice that this procedure does not allow a simple evaluation of the Zak phase as the supercell Hamiltonian generates $2(m^2+n^2)$ bands without well-defined band gaps over the whole BZ, even in the fully 2D case. It is therefore important to calculate the Zak phase from the original Hamiltonian respecting the primitive cell of the lattice, as described in the main text.

\section{Zak phase for arbitrary edges}

\begin{figure}[h]
    \centering
    \includegraphics[width=0.6\linewidth]{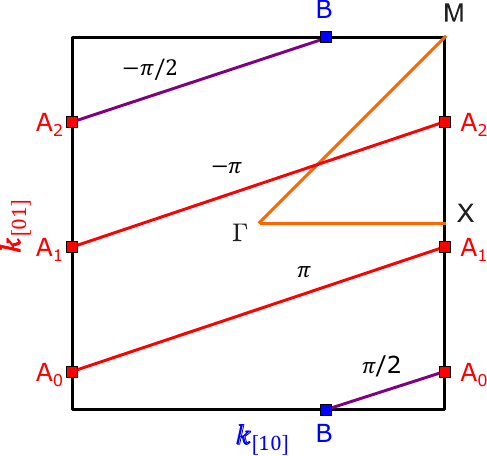}
    \caption{An example of an integration path in the first BZ corresponding to an edge cut along [$\bar{1}$3] direction. The segments of the integration path and their respective contributions to the Zak phase are marked.}
\label{fig:path}
\end{figure}

In the main text we state the geometrical rule assessing whether an edge termination of a particular crystallographic direction supports edge-localized states or not. This rule is formulated for the unperturbed Hamiltonian [Eq.\,\eqref{eq:hamiltonianK}] featuring two quadratic degeneracies, one at $\Gamma$ and one at M. In particular, for an edge defined by a real-space translational vector [$\bar{n}m$] (with $m$ and $n$ integers) the Zak phase must be evaluated along the perpendicular direction [$mn$] in the reciprocal space. Such a path is guaranteed to be closed. An example of such path is shown in Fig.\,\ref{fig:path} for an edge cut along the [$\bar{1}$3] direction. By visualizing the situation using the full reciprocal lattice, we see that the integration path crosses the edges equivalent [10] and [01] BZ boundaries $|n|$ and $|m|$ times, respectively. We exclude the paths that may cross $\Gamma$ or M as the Zak phase is not well defined in such cases. All the crossings of BZ boundaries will split the integration path into unperturbed segments of two kinds. It follows that for $|m|\neq|n|$ there must be segments of the path that directly connect neighboring equivalent BZ boundaries without crossing the other non-equivalent boundaries in between (red segments in Fig.\,\ref{fig:path}). These segments accumulate the Zak phase of $\pm\pi$, i.e., $\pi$ (mod $2\pi$). There will be $||m|-|n||$ such segments, which follows from the unequal number of crossings of the two non-equivalent families of BZ-boundaries. The remaining segments will connect the non-equivalent boundaries, but since the full path is connecting two equivalent points, for every connection of non-equivalent BZ boundaries (cutting the "corner" of the 1$^{\rm st}$ BZ; purple segments in Fig.\,\ref{fig:path}) there exists a complementary connection of non-equivalent boundaries. The complementarity guarantees that the Zak phase accumulated along these corner-cutting segments cancels pairwise. Finally, we see that the Zak phase is therefore determined by the sum over the $N_{\pi}=||m|-|n||$ segments. For odd $N_{\pi}$ the accumulated phase is guaranteed to be non-trivial. For $N_{\pi}$ even, each segment contributes the same $\pi$ (mod $2\pi$) phase and hence adds up to the trivial phase $\gamma_{\rm Zak}=0$ (mod $2\pi$). This leads to the rule $\gamma_{\rm Zak}=N_\pi\pi$ (mod $2\pi$) stated in the main text.  

\section{Pseudospin vs. polarization}

\begin{figure}[h]
    \centering
    \includegraphics[width=0.7\linewidth]{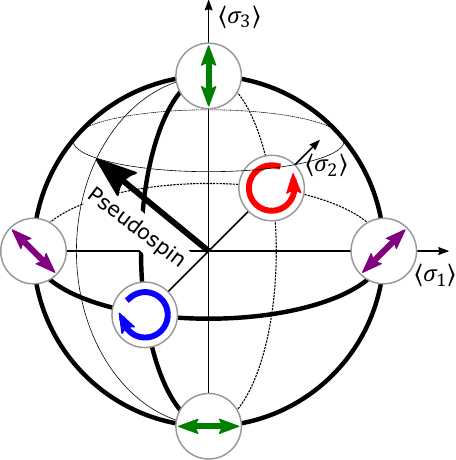}
    \caption{The Bloch sphere defining the polarization state with the two-dimensional representation of the state plotted for the eigenstates of the Pauli operator using polarization ellipses (tracking the head of the vector of polarization as it evolves in time).}
\label{fig:bloch}
\end{figure}

In the main text we distinguish between the concept of pseudospin (closely related to the Stokes vector known in optics \cite{Bohren1998, Saleh1991}), a three-component vector defining the polarization state of the system by the triplet of expectation values of the Pauli spin operators $\sigma_1$, $\sigma_2$ and $\sigma_3$, and the polarization vector  (related to the Jones vector in optics \cite{Saleh1991}), a two-component vector defining the orientation and phase of the oscillation of the two components of the transition dipole moment of the molecule. For clarity, in Fig.\,\ref{fig:bloch} we show the Bloch sphere defining the possible values of the polarization state. The three axes are defined by the expectation values of the respective Pauli matrices. We mark the eigenstates representing the intersections of the sphere with the axes in terms of the polarization ellipses. These have the form of straight lines when the light is linearly polarized. However, in contrast to the two-dimensional vector of polarization, the line marking the polarizaition state does not have a well defined sense. There is therefore a difference in tracking the rotation of a polarization vector and the rotation of the polarization-state line: by performing a half-rotation of the vector we actually complete a full rotation of the polarization line. This is an important distinction when considering pseudospin winding and evaluating the Berry phase.

%


\end{document}